\documentclass[twocolumn,aps,tightenlinefs,superscriptaddress,preprintnumbers]{revtex4}
\usepackage{xparse}
\usepackage{graphicx,color,dcolumn,booktabs,bm}
\usepackage{longtable,lscape}
\usepackage{amsmath}
\usepackage{indentfirst}
\usepackage{epsfig}
\usepackage{feynmf}   
\usepackage{epstopdf}   
\usepackage{slashed}  
\usepackage{cases}
\usepackage{color}
\usepackage{multirow}
\usepackage{ulem}
\usepackage{verbatim}
\usepackage[colorlinks,linkcolor=red,anchorcolor=green,citecolor=blue]{hyperref}
\usepackage{mathrsfs}
\usepackage{rotating}
\usepackage{threeparttable}
\usepackage{subfigure}

\graphicspath{{ps}}

\newcommand{\Dsp}{\ensuremath{D_s^+}}

\newcommand{\Dssp}{\ensuremath{D_s^{*+}}}
\newcommand{\Dssm}{\ensuremath{D_s^{*-}}}

\newcommand{\Dsz}{\ensuremath{D_{s0}^*(2317)}}
\newcommand{\Dszp}{\ensuremath{D_{s0}^*(2317)^+}}
\newcommand{\Dszm}{\ensuremath{D_{s0}^*(2317)^-}}

\newcommand{\Dso}{\ensuremath{D_{s1}(2460)}}
\newcommand{\Dsop}{\ensuremath{D_{s1}(2460)^+}}
\newcommand{\Dsom}{\ensuremath{D_{s1}(2460)^-}}

\newcommand{\Dsoop}{\ensuremath{D_{s1}(2536)^+}}
\newcommand{\Dsoom}{\ensuremath{D_{s1}(2536)^-}}

\newcommand{\Dsjm}{\ensuremath{D_{sJ}^-}}

\newcommand{\ee}{\ensuremath{e^+e^-}}

\begin{document}



\title{ Measurements of Born Cross Sections of $e^+e^-\to D_s^{*+} D_{sJ}^{-} +c.c.$}

\author{
M.~Ablikim$^{1}$, M.~N.~Achasov$^{10,b}$, P.~Adlarson$^{67}$, S. ~Ahmed$^{15}$, M.~Albrecht$^{4}$, R.~Aliberti$^{28}$, A.~Amoroso$^{66A,66C}$, M.~R.~An$^{32}$, Q.~An$^{63,49}$, X.~H.~Bai$^{57}$
% ...
\\
\vspace{0.2cm}
(BESIII Collaboration)\\
\vspace{0.2cm} {\it
$^{1}$ Institute of High Energy Physics, Beijing 100049, People's Republic of China\\
% ...
$^{4}$ Bochum Ruhr-University, D-44780 Bochum, Germany\\
% ...
$^{10}$ G.I. Budker Institute of Nuclear Physics SB RAS (BINP), Novosibirsk 630090, Russia\\
% ...
$^{15}$ Helmholtz Institute Mainz, Staudinger Weg 18, D-55099 Mainz, Germany\\
% ...
$^{28}$ Johannes Gutenberg University of Mainz, Johann-Joachim-Becher-Weg 45, D-55099 Mainz, Germany\\
% ...
$^{32}$ Liaoning Normal University, Dalian 116029, People's Republic of China\\
% ...
$^{49}$ State Key Laboratory of Particle Detection and Electronics, Beijing 100049, Hefei 230026, People's Republic of China\\
% ...
$^{57}$ University of Jinan, Jinan 250022, People's Republic of China\\
% ...
$^{63}$ University of Science and Technology of China, Hefei 230026, People's Republic of China\\
% ...
$^{66}$ University of Turin and INFN, (A)University of Turin, I-10125, Turin, Italy; (B)University of Eastern Piedmont, I-15121, Alessandria, Italy; (C)INFN, I-10125, Turin, Italy\\
$^{67}$ Uppsala University, Box 516, SE-75120 Uppsala, Sweden\\
% ...
$^{b}$ Also at the Novosibirsk State University, Novosibirsk, 630090, Russia\\
% ...
}
}


\begin{abstract}
  The Born cross sections are measured for the first time for the processes $e^+e^-\to D_s^{*+}D_{s0}^*(2317)^- +c.c.$ and $e^+e^-\to D_s^{*+}D_{s1}(2460)^- +c.c.$ at the center-of-mass energy $\sqrt{s}=$ 4.600~GeV, 4.612~GeV, 4.626~GeV, 4.640~GeV, 4.660~GeV, 4.68~GeV, and 4.700~GeV, and for $e^+e^-\to D_s^{*+}D_{s1}(2536)^- +c.c.$ at $\sqrt{s}=$ 4.660~GeV, 4.680~GeV, and 4.700~GeV, using data samples collected with the BESIII detector at the BEPCII collider.   No structures are observed in   cross-section distributions for any of the processes.
\end{abstract}

\maketitle

\tighten

\section{\boldmath Introduction}

The $D_s^+$ meson and its excited states are formed from $c\bar{s}$ quark-antiquark pairs.
Throughout the paper, charge-conjugation is implied.
Three excited $P$-wave states above the $D^{(*)}K$ threshold have been observed at the CLEO, BABAR, Belle, and BESIII experiments~\cite{Ds0(2317)CLEO2003,Ds1(2536)CLEO,Ds0(2317)BaBar2003,Ds1(2536)BaBar,Ds0(2317)BaBar2004,Ds1(2460)Babar2004,Ds0(2317)Belle2003,Ds1(2460)Belle2003,DsDKBES}. They are referred to as \Dszp{}, \Dsop{}, and \Dsoop{}, and are assigned the spin-parity quantum numbers $J^P$ as $0^+$, $1^+$, and $1^+$, respectively~\cite{PDG2020}, matching the predictions from the heavy-quark effective theory~\cite{potModel1985,potModel1995}. Their masses are measured to be $(2317.8\pm 0.5)$~MeV$/c^2$, $(2459.5\pm 0.6)$~MeV$/c^2$, and $(2535.11\pm 0.06)$~MeV$/c^2$, respectively~\cite{PDG2020}.  For the cases of the \Dsz$^+$~and \Dso$^+$ states, these values are significantly lower than the theoretical predictions for the charmed-strange
mesons in the $P$-wave doublet~\cite{chenhx}. The low-mass puzzle of the \Dszp{}~and \Dsop{} mesons has inspired
various exotic explanations, including tetraquark states~\cite{tetra1,tetra2,tetra3,tetra4,tetra5,tetra6}, $D^{(*)}K$ molecule states~\cite{mole1,mole2,mole3,mole4,mole5}, or mixtures of $c\bar{s}$ and $D^{(*)}K$ states~\cite{mix1}. Further experimental measurements are needed in order to elucidate their structures.

Moreover, several charmonium-like $Y$ states with $J^{PC}=1^{--}$ lying above the open-charm threshold have been discovered, such as the $Y(4260)$~\cite{Y42601,Y42602,Y42603}, $Y(4360)$~\cite{Y4360,Y4660}, and $Y(4660)$~\cite{Y4660}. Since $Y$ states could decay to open-charm meson pairs, exclusive cross section measurements of open-charm meson pair production in $e^+ e^-$ collisions will provide further insight on the internal structures of these charmonium-like resonances. Measurements of $\ee\to D_s^{(*)+}D_s^{(*)-}$ cross sections were performed at Belle~\cite{DsDsBelle}, BABAR~\cite{DsDsBaBar}, and CLEO~\cite{DsDsCLEO}. BESIII extended these studies to higher excited states by measuring the cross sections of $\ee\to\Dsp\Dsom$ and $\ee\to\Dssp\Dsom$~\cite{DsDs1BES}  with an integrated luminosity of 859~$\mathrm{pb}^{-1}$ at center-of-mass (c.m.) energies $\sqrt{s}$ from 4.467 to 4.600~GeV. 
In this paper, we report new measurements of the Born cross sections of  the processes $\ee\to\Dssp\Dszm$, $\ee\to\Dssp\Dsom$, and $\ee\to\Dssp\Dsoom$ at BESIII at $\sqrt{s}\geq$ 4.6~GeV, and search for possible vector charmonium-like states.

\section{\boldmath Detector, Data Samples and Monte Carlo Simulations}

The BESIII detector located at the Beijing Electron Positron Collider (BEPCII)~\cite{besiii1} is a major upgrade of the BESII experiment at the BEPC accelerator~\cite{besiii2}. The cylindrical core of the BESIII detector consists of a helium-based multilayer drift chamber (MDC), a plastic scintillator time-of-flight system (TOF), and a CsI(Tl) electromagnetic calorimeter (EMC), which are all enclosed in a superconducting solenoidal magnet providing a 1.0 T magnetic field. The solenoid is supported by an octagonal flux-return yoke with resistive plate counter muon-identifier modules interleaved with steel. The acceptance of charged particles and photons is 93\% over the 4$\pi$ solid angle. The charged-particle momentum resolution at 1~GeV/$c$ is 0.5\%, and the $\mathrm{d}E/\mathrm{d}x$ resolution is 6\% for the electrons from Bhabha scattering. The EMC measures photon energies with a resolution of 2.5\% (5\%) at 1~GeV in the barrel (end-cap) region. The time resolution of the TOF barrel part is 68 ps. The time resolution of the end cap part was
110 ps before 2015; at that time the end cap TOF system was upgraded with multi-gap resistive plate chambers, and the time resolution improved to 60 ps.

The results reported in this article for the processes $\ee\to\Dssp\Dszm$ and $\ee\to\Dssp\Dsom$ are determined from seven energy points from 4.600 to 4.700~GeV, where the data at 4.600~GeV were accumulated in 2014, and the remainder in 2020. At energies of  4.660~GeV and above, the data are also used to study $\ee\to\Dssp\Dsoom$. Table~\ref{tab:1} lists the individual c.m.\ energies and their integrated luminosities.

The {\sc GEANT4}-based~\cite{GEANT} Monte Carlo (MC) simulation framework {\sc BOOST}~\cite{boost}, which includes the description of the detector geometry and response, is used to produce large simulated event samples. These samples are exploited to optimize the event selection criteria, to determine the detection efficiency, and to evaluate the initial-state radiation (ISR) correction factor $(1+\delta)$. Exclusive phase space (PHSP) MC samples of $\ee\to\Dssp\Dszm$, $\ee\to\Dssp\Dsom$, and $\ee\to\Dssp\Dsoom$ are generated using {\sc KKMC}~\cite{KKMC,KKMC2,KKMC3}, where the effects of ISR and beam-energy spread are taken into account. Generic MC samples of open-charm processes are used to estimate background contributions. The known decay modes are modeled with {\sc BESEVTGEN}~\cite{besevtgen,besevtgen2}, using branching fractions taken from PDG~\cite{PDG2020}, while the unknown decays of charmonium states are modeled using {\sc LUNDCHARM}~\cite{lundcharm}. Final-state radiation effects are simulated by the {\sc PHOTOS}~\cite{photos} package. In the signal MC samples, \Dssp{} are simulated to decay into $\gamma\Dsp$, with the subsequent decay of $\Dsp$ into $K^+K^-\pi^+$, while the \Dszm, \Dsom, and \Dsoom{} mesons decay inclusively. A $P$-wave model and a Dalitz plot decay model~\cite{dpgen,dpgenBES,dpgenBaBar} are used to simulate $\Dssp\to\gamma\Dsp$ and $\Dsp\to K^+K^-\pi^+$, respectively.

\section{\boldmath Selection Criteria}

The candidate events of $\ee\to\Dssp\Dszm$, $\ee\to\Dssp\Dsom$, and $\ee\to\Dssp\Dsoom$ are selected with a partial reconstruction method to obtain higher efficiencies. The \Dszm, \Dsom, and \Dsoom{} signals are searched for in the recoil-mass spectrum of \Dssp\ candidates. The \Dssp{} candidates are reconstructed via $\Dssp\to\gamma\Dsp$, while the \Dsp{} candidates are reconstructed via $\Dsp\to\phi\pi^+$, $\phi\to K^+K^-$ and $\Dsp\to\overline{K}^{*0}K^+$, $\overline{K}^{*0}\to K^-\pi^+$. Thus, there are three charged tracks from \Dsp{} decays and one additional photon candidate from $\Dssp\to\gamma\Dsp$.

Each track must originate from the interaction point (IP), which means that the distance of the closest approach to the IP of each track is required to be within 10~cm in the beam direction and within 1~cm in the plane perpendicular to the beam direction. Additionally, each track must reside within the active region of the MDC, meaning that its polar angle $\theta$ must satisfy $|\cos\theta|<$ 0.93. The $\mathrm{d}E/\mathrm{d}x$ and TOF information are used to perform particle identification (PID). Pion candidates are required to satisfy $\mathrm{Prob}(\pi)>\mathrm{Prob}(K)$ and $\mathrm{Prob}(\pi)>0.001$, where $\mathrm{Prob}(\pi)$ and $\mathrm{Prob}(K)$ are the PID probabilities for a track to be a pion and kaon, respectively. Kaon candidates are required to satisfy $\mathrm{Prob}(K)>\mathrm{Prob}(\pi)$ and $\mathrm{Prob}(K)>0.001$. With these PID requirements, the probability of misidentifying a $K^+$ as $\pi^+$ or a $\pi^+$ as $K^+$ is below 3\%.

The photon candidates are selected from EMC showers that are not associated with charged tracks. The deposited energy is required to be larger than 25~MeV in the barrel EMC ($\left|\cos\theta\right|<0.8$), or larger than 50~MeV in the end-cap EMC ($0.86<\left|\cos\theta\right|<0.92$). To eliminate the showers from charged particles, the angle between the photon and the extrapolated impact point of any good charged track at the EMC front face must be larger than $20^\circ$. The timing of the shower is required to be within $[50,700]$~ns after the reconstructed event start time to suppress noise and energy deposits unrelated to the event.

To reconstruct \Dssp\ mesons, two kaons, one pion, and one photon candidates are required. All the $K^+K^-\pi^+$ in all combinations of $\gamma K^+K^-\pi^+$, which pass a vertex fit are kept. Mass-constrained fits to the nominal masses of \Dsp{} and \Dssp{} (2C) are performed, and the $\chi^2$ of this fit
is required to be less than 10 to suppress the backgrounds. To select $\Dsp\to\phi\pi^+$, $\phi\to K^+K^-$ and $\Dsp\to\overline{K}^{*0}K^+$, $\overline{K}^{*0}\to K^-\pi^+$ submodes, the invariant masses of $K^+K^-$ and $K^-\pi^+$ are required to satisfy $|M(K^+K^-)-m_{\phi}|<9$~MeV/$c^2$ and $|M(K^-\pi^+)-m_{\overline{K}^{*0}}|<84$~MeV/$c^2$, respectively, where $m_\phi$ ($m_{\overline{K}^{*0}}$) is the nominal mass of the $\phi$ ($\overline{K}^{*0}$) meson taken from the PDG~\cite{PDG2020}.

\section{\boldmath Measurements of the $\ee\to D_s^{*+} D_{sJ}^-$}

\begin{figure*}[htbp!]
  \begin{center}
    \includegraphics[width=.32\textwidth]{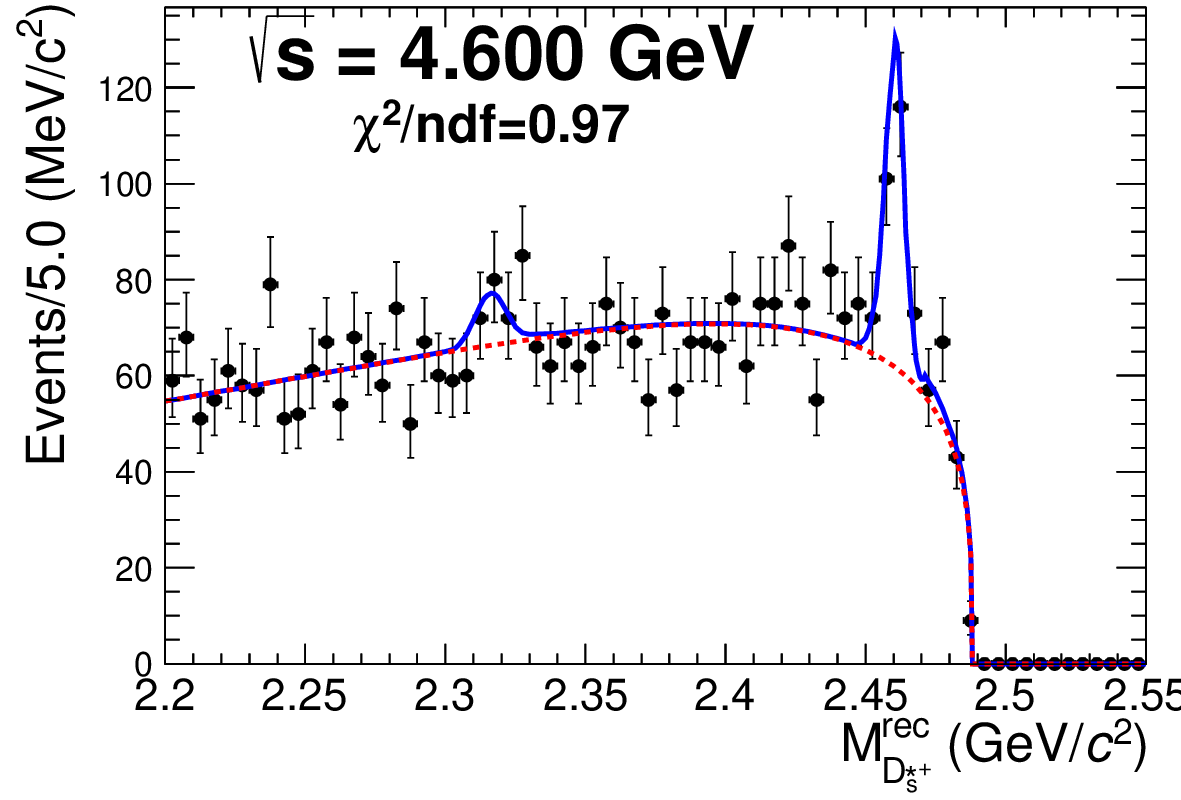}
    \includegraphics[width=.32\textwidth]{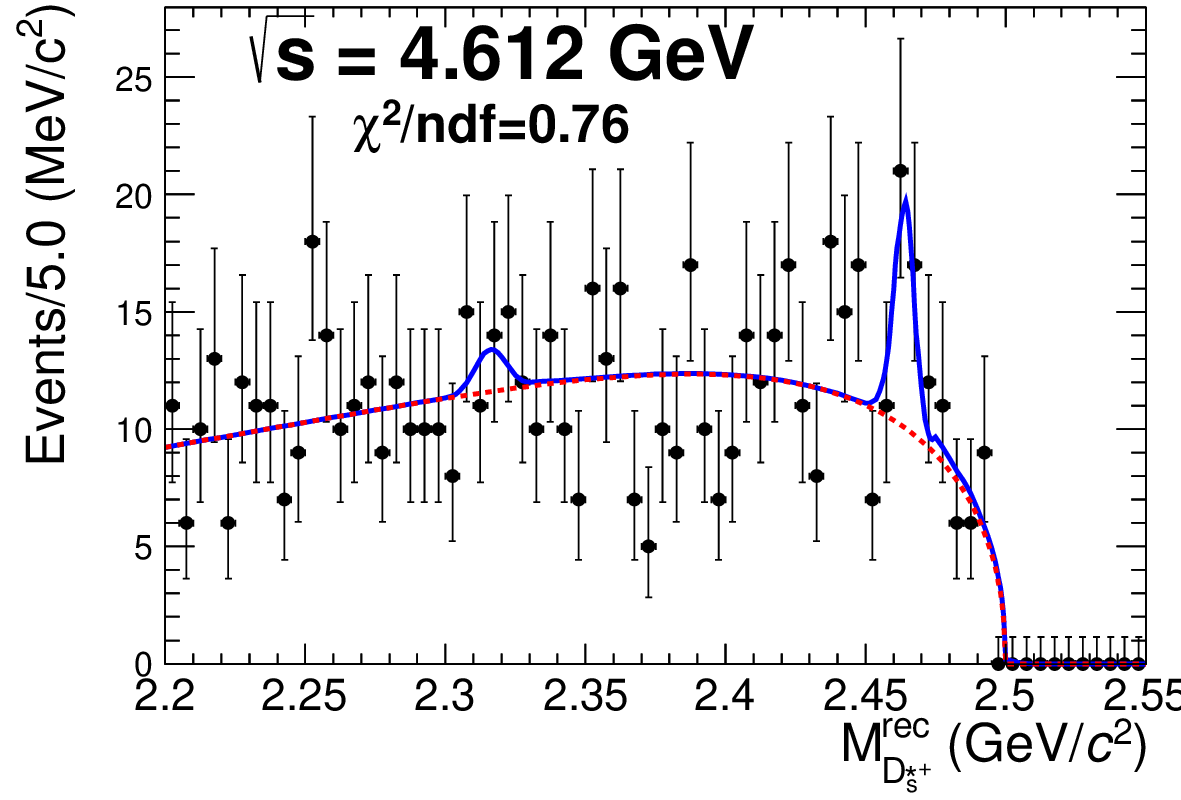}
    \includegraphics[width=.32\textwidth]{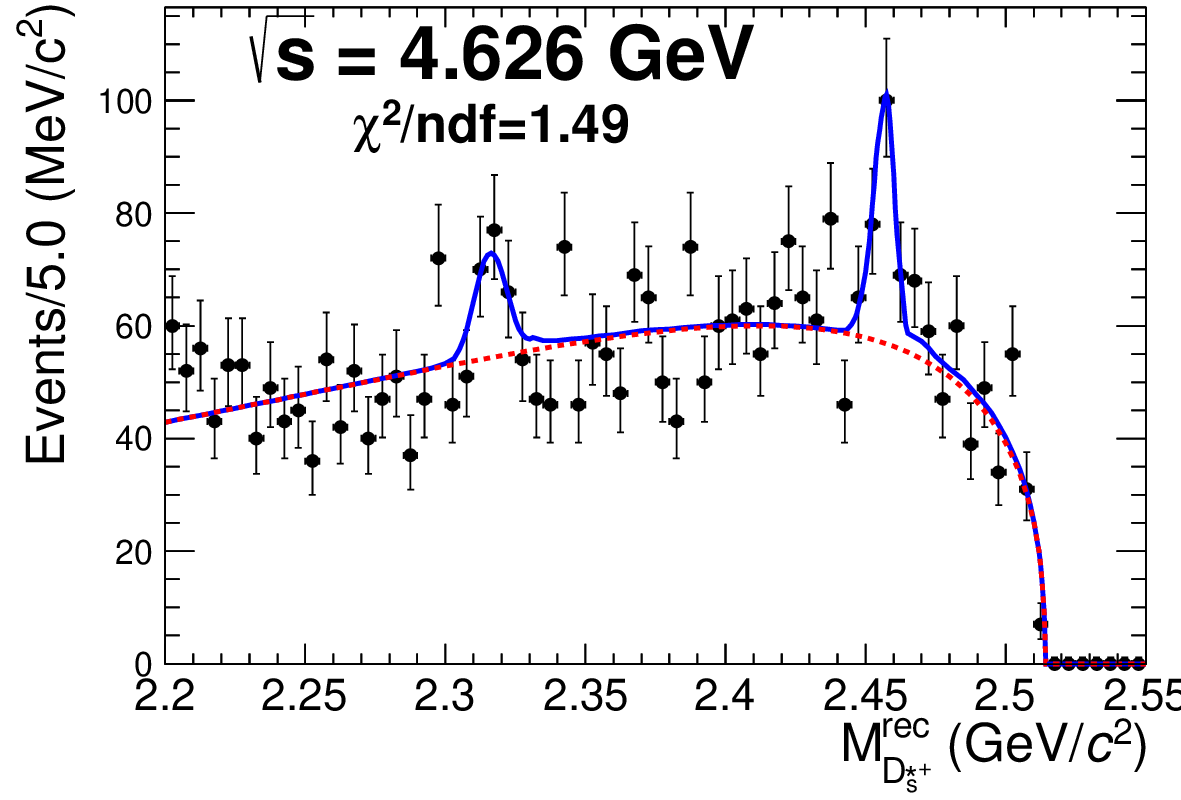}
    \includegraphics[width=.32\textwidth]{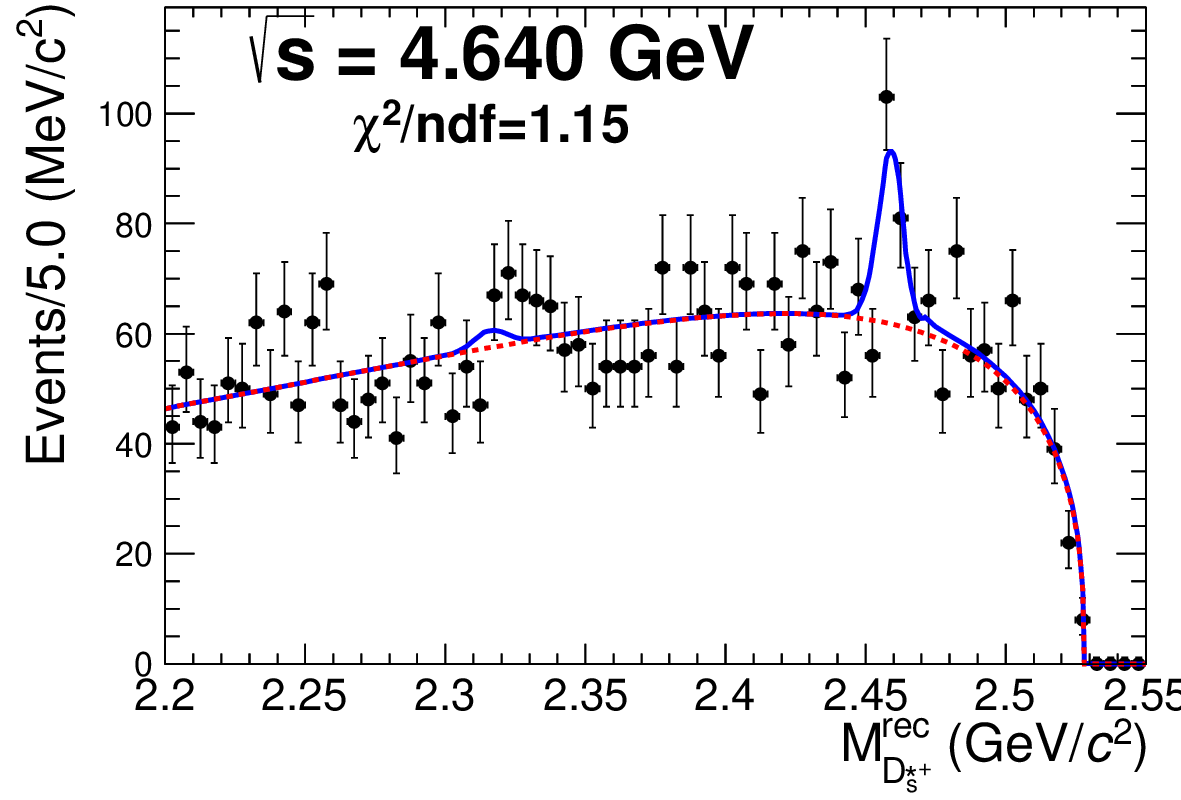}
    \includegraphics[width=.32\textwidth]{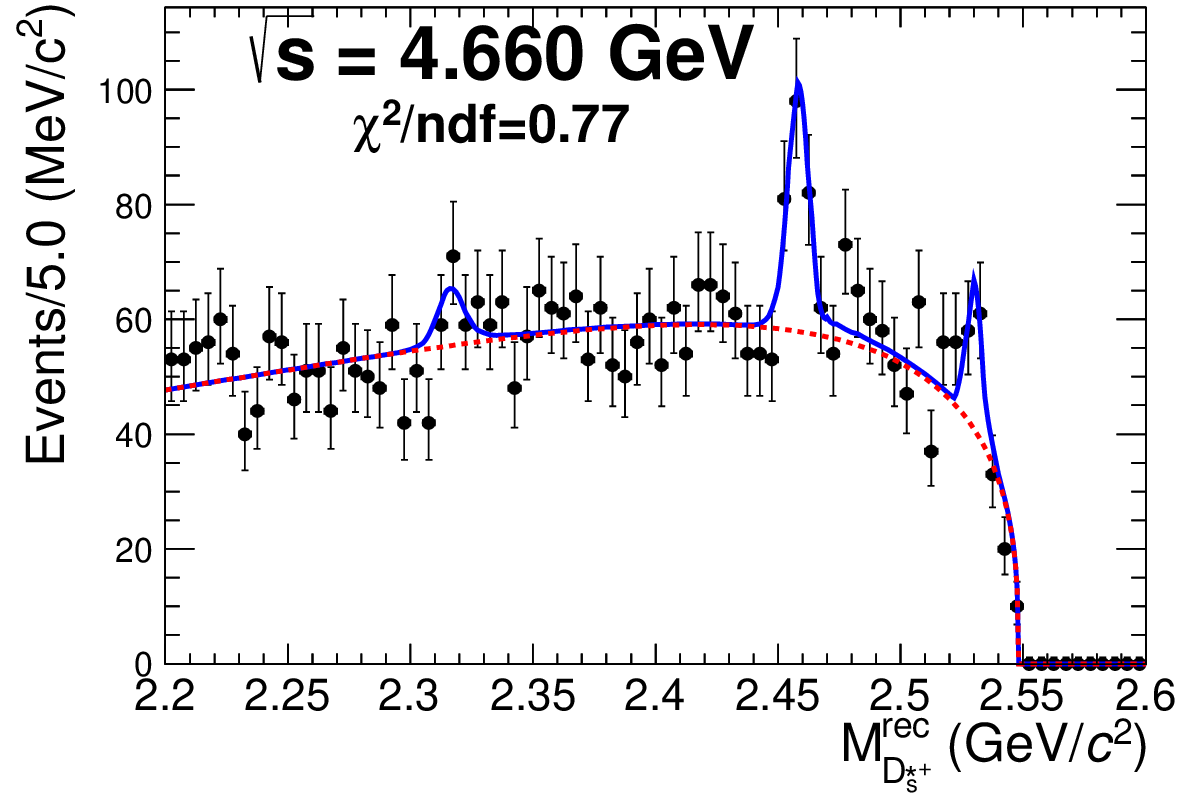}
    \includegraphics[width=.32\textwidth]{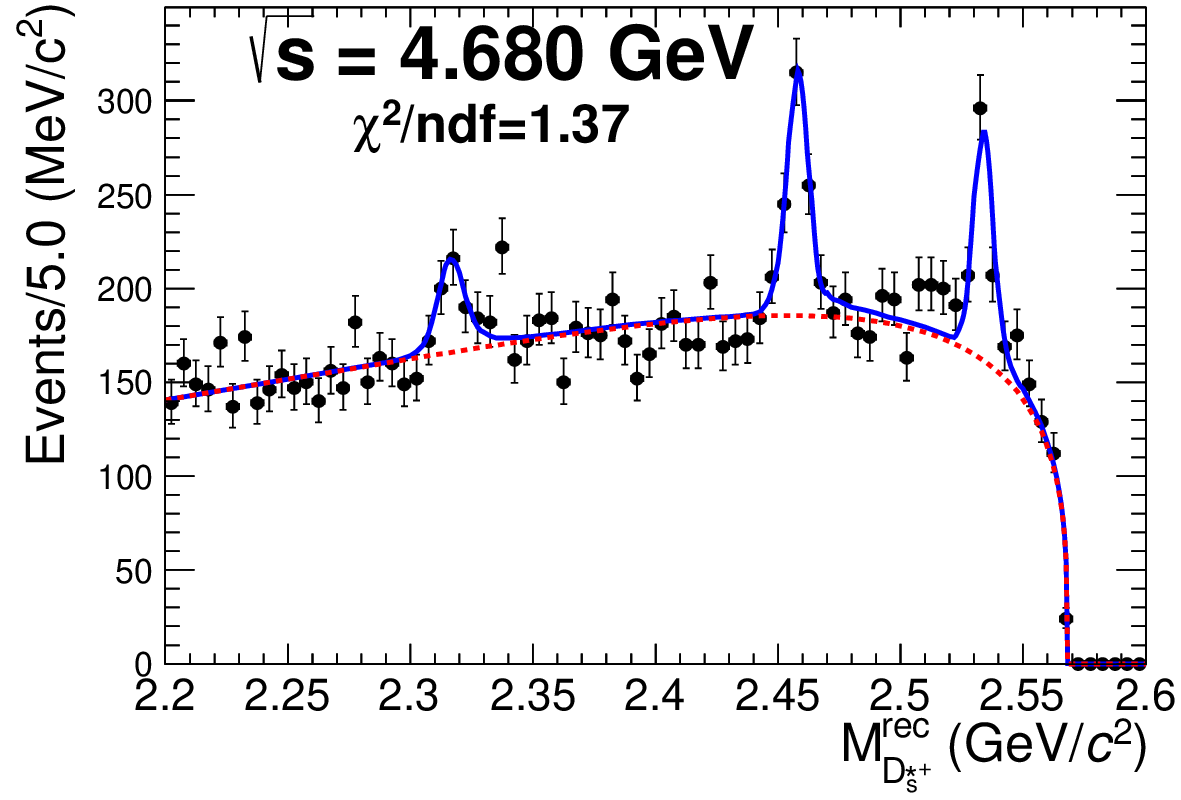}
    \includegraphics[width=.32\textwidth]{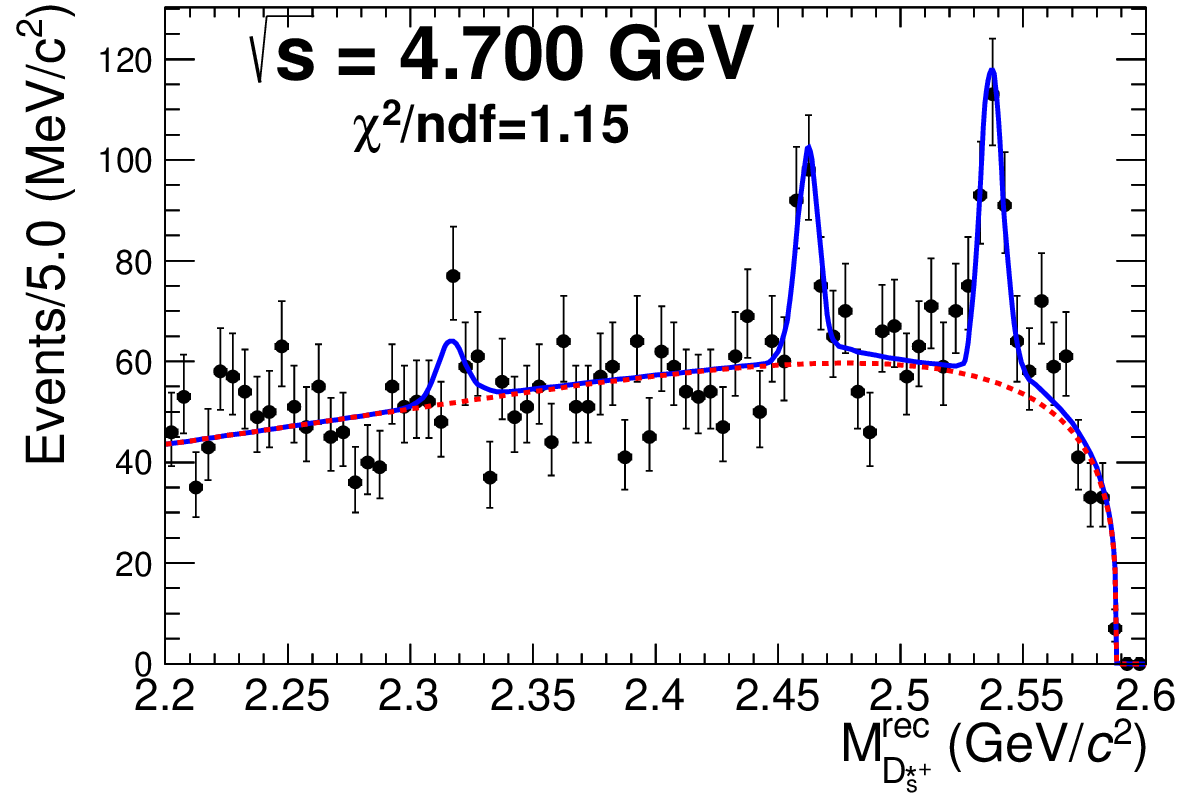}
  \caption{The recoil-mass distributions of \Dssp{} candidates from data samples at $\sqrt{s}=$ 4.600~GeV, 4.612~GeV, 4.626~GeV, 4.640~GeV, 4.660~GeV, 4.680~GeV, and 4.700~GeV, respectively. The dots with error bars are data and the solid lines are the best fits. Clear \Dsom{} signals are observed at all of the energy points except for 4.612~GeV; clear \Dsoom{} signals are observed at $\sqrt{s}=$ 4.660~GeV, 4.680~GeV, and 4.700~GeV; and clear \Dszm{} signals are observed at $\sqrt{s}=$ 4.626~GeV and 4.680~GeV. The fitted results together with the signal significances are summarized in Table~\ref{tab:1}. The fit quality of $\chi^2/n.d.f.$ is presented in each plot, where $n.d.f.$ stands for the number of degrees of freedom.}
  \label{fig:1}
  \end{center}
\end{figure*}

The recoil-mass distributions of \Dssp{} candidates from data samples at $\sqrt{s}=$ 4.600~GeV, 4.612~GeV, 4.626~GeV, 4.640~GeV, 4.660~GeV, 4.680~GeV, and 4.700~GeV are shown in Fig.~\ref{fig:1}.
To improve the recoil mass resolution, the recoil mass of \Dssp{} is defined to be $M_{\Dssp{}}^{\rm rec}=M_{\gamma K^+K^-\pi^+}^{\rm recoil}+M(\gamma K^+K^-\pi^+)-m_{\Dssp{}}$, where $M_{\gamma K^+K^-\pi^+}^{\rm recoil}=\sqrt{\left(P_{\rm c.m.}-P_{\gamma}-P_{K^+}-P_{K^-}-P_{\pi^+}\right)^2}$. Here $P_{\rm c.m.}$, $P_{\gamma}$, $P_{K^+}$, $P_{K^-}$, and $P_{\pi^+}$ are the four-momenta of the initial $\ee$ system, the selected $\gamma$, $K^+$, $K^-$, and $\pi^+$, respectively, $M(\gamma K^+K^-\pi^+)$ is the invariant mass of the $\gamma K^+K^-\pi^+$ system, and $m_{\Dssp{}}$ is the nominal mass of the \Dssp{} meson~\cite{PDG2020}.
Clear signals are seen for the \Dsom{} at all of the energy points apart from 4.612~GeV, and are found for the \Dsoom{} at $\sqrt{s}=$ 4.660~GeV, 4.680~GeV, and 4.700~GeV, and for the \Dszm{} at $\sqrt{s}=$ 4.626~GeV and 4.680~GeV. Detailed studies of the generic MC samples with a generic event type analysis tool, TopoAna~\cite{topo}, indicate that there are no peaking backgrounds in the signal region.
We considered possible contributions of processes such as $e^+e^-\to\Dsp\gamma D_{sJ}^-$ and $e^+e^-\to D^0(\to K^-\pi^+)K^+\gamma\Dszm$ by plotting the recoil mass distributions for events in the \Dssp{} mass sidebands; no peaking structures were observed.
An unbinned maximum-likelihood fit is performed to the \Dssp{}  recoil-mass distributions  to determine the signal yields of \Dsjm\ mesons. The signal probability density function is modeled according to a MC-derived signal shape, while the background is modeled with an ARGUS function~\cite{argus}. The fitted results are summarized in Fig.~\ref{fig:1} and  Table~\ref{tab:1}. The \Dszm{} signal significance of the combined data samples from all energy points is $6.9\sigma$. The signal significances are determined by comparing the log-likelihood values with and without including a \Dsjm{} signal in the fit, taking the change in the number of degrees-of-freedom into account.

Since the statistical significances of \Dsjm{} signals at some energy points are less than $3\sigma$,
with a uniform prior
probability density function,
the Bayesian upper limits on the numbers of \Dsjm{} signal events ($N_{\rm U.L.}$) are determined at the $90\%$ confidence level (C.L.) by solving the following equation
$$\int^{N_{\rm U.L.}}_0 \mathcal{L}(x)dx = 0.9 \int^{+\infty}_0\mathcal{L}(x)dx,$$
\noindent where $x$ is the assumed yield of the \Dsjm{} signal and $\mathcal{L}(x)$ is
the maximized likelihood of the data assuming $x$ signal events. The results of $N_{\rm U.L.}$ obtained by using above method are listed in Table~\ref{tab:1}.

The Born cross sections of $\ee\to\Dssp\Dsjm$ are calculated using the formula
\begin{equation}
\begin{aligned}
\label{sigma-B}
\sigma_{B}(\ee\to\Dssp\Dsjm)=
\frac{N_{\rm fit}}{\mathcal{L}_{\rm int}(1+\delta)(1+\delta^{\rm vp})\epsilon_{\Dssp}},
\end{aligned}
\end{equation}
where $N_{\rm fit}$ is  the fitted \Dsjm{} signal yield, $1+\delta$ is the radiative correction factor obtained from a QED calculation with $1\%$ accuracy~\cite{radiator} using the {\sc KKMC} generator, $1+\delta^{\rm vp}$ is the vacuum polarization factor, which is taken from Ref.~\cite{vacuum_new} ($\delta^{\rm vp}$ is around $0.055$ for all studied energy points), and $\mathcal{L}_{\rm int}$ is the integrated luminosity at each energy point.
The \Dssp{} reconstruction efficiency $\epsilon_{\Dssp}$ is the product of the detection efficiency $\epsilon$ and the branching fractions for $\Dssp\to\gamma\Dsp$ and $\Dsp\to K^+K^-\pi^+$ (93.5\% and 5.39\%, respectively~\cite{PDG2020}).
The calculation of the upper limits on Born cross sections at 90\% C.L. is performed analogously, replacing $N_{\rm fit}$ with $N_{\rm U.L.}$.

The measured Born cross sections of $\ee\to\Dssp\Dsjm$ and the upper limits at 90\% C.L. (with systematic uncertainties included) are summarized in Table~\ref{tab:1}. The systematic uncertainties and the method to take them into account in the upper limits are discussed in Sec.~\ref{sec:sys}. The Born cross sections, with statistical uncertainties only,  are shown in Fig.~\ref{fig:2}.

\begin{table*}[htbp!]
  \begin{center}
  \caption{Summary of the measured Born cross sections for $\ee\to\Dssp\Dszm$, $\ee\to\Dssp\Dsom$, and $\ee\to\Dssp\Dsoom$. Listed in the table are the integrated luminosity $\mathcal{L}_{\rm int}$, the signal efficiency $\epsilon$ from signal MC sample, the number of fitted \Dsjm{} signal events $N_{\rm fit}$, the upper limit at 90\% C.L.\ on the number of fitted \Dsjm{} signal yields $N_{\rm U.L.}$, the ISR radiative correction factor $(1+\delta)$, the statistical signal significance, and the measured Born cross section ($\sigma_B$) and its upper limit ($\sigma_B^{\rm U.L.}$) at 90\% C.L..
  }
  \label{tab:1}
  \begin{tabular}{ccccccD{p}{{}}{5.5}cccccccD{x}{{}}{4.11}}
    \hline
    $\sqrt{s}$ (GeV) && $\mathcal{L}_{\rm int}$ (pb$^{-1}$) && $\epsilon$ (\%) && \multicolumn{1}{c}{$N_{\rm fit}$} && $N_{\rm U.L.}$ && $(1+\delta)$ && significance && \multicolumn{1}{c}{$\sigma_B$ ($\sigma_B^{\rm U.L.}$) (pb)} \\\hline
    \multicolumn{15}{c}{$\ee\to\Dssp\Dszm+c.c.$} \\\hline
    4.600 && $586.89$ && 14.1 && 25.7p^{+19.1}_{-18.3} && 52.0 && 0.821 && $1.4\sigma$ && 6.8x^{+5.0}_{-4.8}~(14.8) \\
    4.612 && $102.50$ && 13.7 && 9.5p^{+8.6}_{-7.8} && 22.0 && 0.826 && $1.2\sigma$ && 14.7x^{+13.3}_{-12.0}~(34.7) \\
    4.626 && $511.06$ && 13.7 && 62.9p~\pm~18.8 && $\cdots$ && 0.830 && $3.6\sigma$ && 19.3x\pm 5.8\pm 2.0 \\
    4.640 && $541.37$ && 13.8 && 20.8p^{+18.1}_{-17.3} && 46.4 && 0.834 && $1.2\sigma$ && 6.0x^{+5.2}_{-5.0}~(14.1) \\
    4.660 && $523.63$ && 13.9 && 20.0p^{+18.2}_{-17.2} && 45.6 && 0.838 && $1.2\sigma$ && 5.8x^{+5.3}_{-5.0}~(14.1) \\
    4.680 && $1643.38$ && 14.0 && 151.9p~\pm~33.2 && $\cdots$ && 0.845 && $4.9\sigma$ && 13.9x\pm 3.0\pm 1.4 \\
    4.700 && $526.20$ && 13.8 && 47.5p^{+19.5}_{-18.6} && 73.6 && 0.849 && $2.7\sigma$ && 13.7x^{+5.6}_{-5.4}~(21.4) \\\hline
    \multicolumn{15}{c}{$\ee\to\Dssp\Dsom+c.c.$} \\\hline
    4.600 && $586.89$ && 14.1 && 107.6p~\pm~17.8 && $\cdots$ && 0.743 && $7.1\sigma$ && 31.2x\pm 5.2\pm 3.7 \\
    4.612 && $102.50$ && 13.7 && 15.8p^{+7.7}_{-7.0} && 26.8 && 0.766 && $2.5\sigma$ && 26.1x^{+12.8}_{-11.5}~(44.4)\\
    4.626 && $511.06$ && 13.5 && 88.2p~\pm~18.3 && $\cdots$ && 0.783 && $5.6\sigma$ && 29.1x\pm 6.0\pm 2.6 \\
    4.640 && $541.37$ && 13.6 && 75.2p~\pm~18.3 && $\cdots$ && 0.796 && $4.7\sigma$ && 22.8x\pm 5.6\pm 2.3 \\
    4.660 && $523.63$ && 13.5 && 100.6p~\pm~19.3 && $\cdots$ && 0.811 && $6.1\sigma$ && 31.1x\pm 6.0\pm 2.7 \\
    4.680 && $1643.38$ && 14.0 && 339.0p~\pm~35.0 && $\cdots$ && 0.822 && $11.0\sigma$ && 31.9x\pm 3.3\pm 2.5 \\
    4.700 && $526.20$ && 13.7 && 103.4p~\pm~20.2 && $\cdots$ && 0.831 && $5.8\sigma$ && 30.8x\pm 6.0\pm 2.6 \\\hline
    \multicolumn{15}{c}{$\ee\to\Dssp\Dsoom+c.c.$} \\\hline
    4.660 && $523.63$ && 13.8 && 35.0p~\pm~11.5 && $\cdots$ && 0.639 && $3.4\sigma$ && 13.4x\pm 4.4\pm 1.7 \\
    4.680 && $1643.38$ && 13.9 && 243.7p~\pm~27.9 && $\cdots$ && 0.706 && $10.1\sigma$ && 26.9x\pm 3.1\pm 2.3 \\
    4.700 && $526.20$ && 14.0 && 109.7p~\pm~18.7 && $\cdots$ && 0.753 && $7.0\sigma$ && 35.1x\pm 6.0\pm 3.0 \\\hline
  \end{tabular}
  \end{center}
\end{table*}

\section{\boldmath Systematic Uncertainties}
\label{sec:sys}

The systematic uncertainties of the measured cross sections of $\ee\to\Dssp\Dszm$, $\ee\to\Dssp\Dsom$, and $\ee\to\Dssp\Dsoom$ are divided into two categories: multiplicative systematic uncertainties and additive systematic uncertainties. The multiplicative systematic uncertainties are associated with  tracking and PID efficiencies, photon-detection efficiency, MC sample size, ISR and vacuum-polarization corrections, measurement of luminosity, the branching fractions of intermediate states, and the kinematic fit. The additive systematic uncertainties are associated with the \Dsjm{} mass, the background shape, and the fit range.

The uncertainties related to PID and tracking are determined to be 3.0\% respectively~\cite{trackingZcs}. The uncertainty of the photon reconstruction efficiency is 1.0\%, which is derived from the study of $J/\psi\to\rho^0(\to\pi^+\pi^-)\pi^0(\to\gamma\gamma)$~\cite{photonpi0} events. The uncertainties due to finite sizes of the MC samples are determined to be at most 0.9\% at each energy point, which arises from the statistical uncertainty of the selection efficiency measured from these samples. From an MC study, we find that the systematic uncertainty due to the choice of $\Dsp\to K^+K^-\pi^+$ decay model is negligible. The shapes of the cross section of the processes $e^+e^- \to\Dssp\Dsjm$ affect the radiative correction factor and the detection efficiency. Due to the limited number of events in the data sample, a detailed determination of the energy dependence (``line shape''), which would allow for an iterative determination of radiative correction factors, is not possible. Therefore, the input line shapes are changed to a first order polynomial multiplied by
$\sqrt{E_m-E_0}$, a function that reasonably describes the shape of available data, where $E_m$ stands for the c.m.\ energy and $E_0$ stands for the threshold energy for each process, and the differences in the selection efficiency $\varepsilon(1+\delta)$ are taken as the systematic uncertainties. The uncertainty from vacuum polarization is less than 0.1\%~\cite{vacuum_new}, which is negligible compared to other sources of uncertainties. The integrated luminosities of the data samples are measured using large-angle Bhabha scattering events with an uncertainty less than 1.0\%. The uncertainties in the branching fractions $\mathcal{B}(\Dsp\to K^+K^-\pi^+)$ and $\mathcal{B}(\Dssp\to\gamma\Dsp)$ are 2.8\% and 0.7\%~\cite{PDG2020}, respectively. The uncertainty of the 2C kinematic fit is estimated using control samples of $\ee\to\Dssp\Dssm$ events at $\sqrt{s}=$ 4.42~GeV and 4.6~GeV. The difference between the data and MC efficiencies due to the 2C kinematic fit is 1.7\%, which is taken as the systematic uncertainty. Using a control sample of $e^+e^-\to\Dssp\Dssm$, the mass resolutions of $\Dssm$ candidates between signal MC simulation and data are consistent. Thus, the systematic uncertainty due to mass resolution is negligible.

The uncertainties due to \Dsjm{} mass are estimated by varying its value by the measured uncertainty~\cite{PDG2020}. The differences in the fitted \Dsjm{} signal yields are taken as the systematic uncertainties. Using a control sample of $e^+e^-\to\Dssp\Dssm$ decays, we find the resolution in missing mass to be essentially the same in data as in the MC, at the current level of statistics. The systematic uncertainties due to fitting itself, the background shape, and the fit range are estimated with a toy MC method. We simulate an ensemble of experiments, generating \Dssp{} recoil mass distributions based on the nominal fitted results. We generate 1000 distributions and subsequently fit them to obtain \Dsjm{} signal yields. We plot these results in Gaussian distributions. The difference between the mean values of these distributions and the input signal yields represents the systematic bias or uncertainty due to the fitting procedure. The uncertainty is found to be negligible for all \Dsjm{} signals.
A similar method is used to estimate the systematic uncertainties due to the fit range and the background shape. For the fit range, the lower bound is changed from 2.20~GeV$/c^2$ to 2.18~GeV$/c^2$ and to 2.22~GeV$/c^2$. For the background shape, instead of an ARGUS function~\cite{argus}, a polynomial $f(M)=(M-M_a)^c(M_b-M)^d$ is used, where $M_a$ and $M_b$ are the lower and upper thresholds of the \Dssp{} recoil mass distribution.


For those energy points with a \Dsjm{} statistical significance larger than 3$\sigma$, the central values of the cross section with statistical and systematic uncertainties are reported, and the systematic uncertainties are summarized in Table~\ref{tab:2}. For the other energy points, the upper limits on the cross section at 90\% C.L. are reported and the systematic uncertainties are taken into account in two steps. Firstly, among the additive systematic uncertainties described above, the most conservative upper limit at 90\% C.L.\ is kept. Then, to take into account the multiplicative systematic uncertainty, the likelihood curve is convolved with a Gaussian function with a width parameter equal to the corresponding total multiplicative systematic uncertainty. All of the multiplicative systematic uncertainties for the energy points with a \Dsjm{} signal significance less than $3\sigma$ are summarized in Table~\ref{tab:3}. Assuming that all the sources are independent, the total systematic uncertainty is obtained by adding them in quadrature. The final results of the Born cross section with systematic uncertainties considered are listed in Table~\ref{tab:1}, and shown in Fig.~\ref{fig:2} with statistical error bars only.

\begin{table*}[htbp!]
  \centering
  \caption{Summary of the systematic uncertainties (\%) in $\sigma_B(\ee\to\Dssp\Dsjm)$ for those energy points with statistical significances larger than $3\sigma$.}
  \label{tab:2}
    \begin{tabular}{cccp{0.15cm}ccccccp{0.15cm}ccc}
      \hline
      \multirow{2}{*}{Sources $/\sqrt{s}$ (GeV)} & \multicolumn{2}{c}{$\Dszm$} && \multicolumn{6}{c}{$\Dsom$} && \multicolumn{3}{c}{$\Dsoom$} \\\cline{2-14}
      & 4.626 & 4.680 && 4.600 & 4.626 & 4.640 & 4.660 & 4.680 & 4.700 && 4.660 & 4.680 & 4.700 \\\hline
      Tracking, PID and photon detection & 7.0 & 7.0 && 7.0 & 7.0 & 7.0 & 7.0 & 7.0 & 7.0 && 7.0 & 7.0 & 7.0 \\
      MC statistics & 0.9 & 0.8 && 0.8 & 0.9 & 0.9 & 0.9 & 0.8 & 0.9 && 0.9 & 0.8 & 0.8 \\
      Kinematic fit & 1.7 & 1.7 && 1.7 & 1.7 & 1.7 & 1.7 & 1.7 & 1.7 && 1.7 & 1.7 & 1.7 \\
      \Dsjm{} mass & 5.1 & 3.9 && 1.5 & 1.7 & 3.0 & 2.3 & 0.7 & 2.2 && 4.7 & 2.1 & 1.3 \\
      Fit range & 3.2 & 2.9 && 6.2 & 0.4 & 2.1 & 2.6 & 0.9 & 1.5 && 8.1 & 1.5 & 1.5 \\
      Background shape & 2.4 & 2.4 && 5.2 & 3.2 & 4.5 & 0.1 & 0.2 & 1.5 && 2.3 & 1.6 & 2.4 \\
      ISR radiative correction & 2.5 & 2.7 && 3.2 & 2.3 & 1.8 & 1.0 & 0.1 & 0.9 && 3.6 & 1.9 & 0.8 \\
      Luminosity & 1.0 & 1.0 && 1.0 & 1.0 & 1.0 & 1.0 & 1.0 & 1.0 && 1.0 & 1.0 & 1.0 \\
      Branching fraction & 2.8 & 2.8 && 2.8 & 2.8 & 2.8 & 2.8 & 2.8 & 2.8 && 2.8 & 2.8 & 2.8 \\\hline
      Total & 10.5 & 9.9 && 11.8 & 9.0 & 9.9 & 8.6 & 7.9 & 8.5 && 13.0 & 8.6 & 8.5 \\\hline
    \end{tabular}
\end{table*}

\begin{table*}[htbp!]
  \centering
    \caption{Summary of the multiplicative systematic uncertainties (\%) in $\sigma_B(\ee\to\Dssp\Dsjm)$ for those energy points with statistical significances less than $3\sigma$.}
  \label{tab:3}
    \begin{tabular}{ccccccp{0.15cm}c}
      \hline
      \multirow{2}{*}{Sources $/\sqrt{s}$ (GeV)} & \multicolumn{5}{c}{$\Dszm$} && $\Dsom$ \\\cline{2-8}
      & 4.600 & 4.612 & 4.640 & 4.660 & 4.700 && 4.612 \\\hline
      Tracking, PID and photon detection & 7.0 & 7.0 & 7.0 & 7.0 & 7.0 && 7.0 \\
      MC statistics & 0.8 & 0.9 & 0.9 & 0.8 & 0.9 && 0.9 \\
      Kinematic fit & 1.7 & 1.7 & 1.7 & 1.7 & 1.7 && 1.7 \\
      ISR radiative correction & 2.5 & 7.3 & 2.8 & 2.4 & 2.5 && 5.0 \\
      Luminosity & 1.0 & 1.0 & 1.0 & 1.0 & 1.0 && 1.0 \\
      Branching fraction & 2.8 & 2.8 & 2.8 & 2.8 & 2.8 && 2.8 \\\hline
      Total & 8.3 & 10.7 & 8.4 & 8.2 & 8.3 && 9.3 \\\hline
    \end{tabular}
\end{table*}

\section{\boldmath Summary}

\begin{figure}[htbp!]
  \begin{center}
  \includegraphics[width=.4\textwidth]{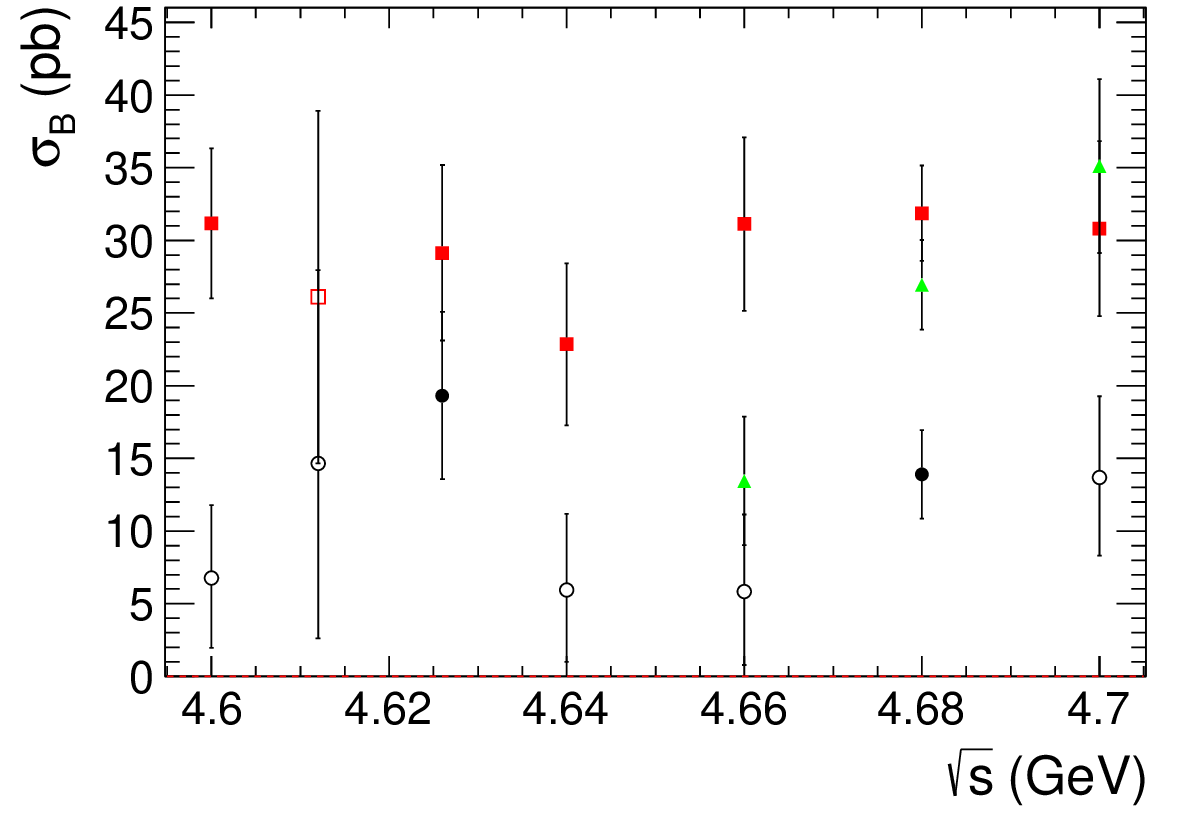}
  \caption{The Born cross sections of $\ee\to\Dssp\Dszm$ (the dots with error bars), $\ee\to\Dssp\Dsom$ (the rectangles with error bars), and $\ee\to\Dssp\Dsoom$ (the triangles with error bars) from 4.6~GeV to 4.7~GeV, where the error bars are statistical only. The energy points with signal significances less than $3\sigma$ are marked empty, and the upper limits at 90\% C.L.\ on their cross sections are given in Table~\ref{tab:1}.}
  \label{fig:2}
  \end{center}
\end{figure}

In summary, signals are observed in the $\ee\to\Dssp\Dszm$ process from $e^+ e^-$ collision data samples at c.m.\ energies of 4.626~GeV and 4.68~GeV, in the $\ee\to\Dssp\Dsom$ process from data samples at c.m.\ energies of 4.6~GeV, 4.626~GeV, 4.64~GeV, 4.66~GeV, 4.68~GeV, and 4.7~GeV, and the $\ee\to\Dssp\Dsoom$ process from data samples at c.m.\ energies of 4.66~GeV, 4.68~GeV, and 4.7~GeV, all with statistical significances larger than $3\sigma$. The Born cross sections of these processes are measured for the first time. The results are listed in Table~\ref{tab:1}, and shown in Fig.~\ref{fig:2} with statistical error bars only. No significant structures are observed in the measured cross sections.

\section{\boldmath Acknowledgement}

The BESIII collaboration thanks the staff of BEPCII and the IHEP computing center for their strong support. This work is supported in part by National Key R\&D Program of China under Contracts Nos. 2020YFA0406300, 2020YFA0406400; National Natural Science Foundation of China (NSFC) under Contracts Nos. 11625523, 11635010, 11735014, 11822506, 11835012, 11935015, 11935016, 11935018, 11961141012, 12022510, 12025502, 12035009, 12035013, 12061131003; the Chinese Academy of Sciences (CAS) Large-Scale Scientific Facility Program; Joint Large-Scale Scientific Facility Funds of the NSFC and CAS under Contracts Nos. U1732263, U1832207; CAS Key Research Program of Frontier Sciences under Contract No. QYZDJ-SSW-SLH040; 100 Talents Program of CAS; INPAC and Shanghai Key Laboratory for Particle Physics and Cosmology; ERC under Contract No. 758462; European Union Horizon 2020 research and innovation programme under Contract No. Marie Sklodowska-Curie grant agreement No 894790; German Research Foundation DFG under Contracts Nos. 443159800, Collaborative Research Center CRC 1044, FOR 2359, FOR 2359, GRK 214; Istituto Nazionale di Fisica Nucleare, Italy; Ministry of Development of Turkey under Contract No. DPT2006K-120470; National Science and Technology fund; Olle Engkvist Foundation under Contract No. 200-0605; STFC (United Kingdom); The Knut and Alice Wallenberg Foundation (Sweden) under Contract No. 2016.0157; The Royal Society, UK under Contracts Nos. DH140054, DH160214; The Swedish Research Council; U. S. Department of Energy under Contracts Nos. DE-FG02-05ER41374, DE-SC-0012069

\end{document}